\newcommand{\singlefig}[2]{
\vspace{0.5cm}
\begin{center}
\begin{minipage}{#1}
\epsfxsize=#1
\epsffile{#2}
\end{minipage}
\end{center}}

\newenvironment{figcaption}[2]{
 \vspace{0.3cm}
 \refstepcounter{figure}
 \label{#1}
 \begin{center}
 \begin{minipage}{#2}
 \begingroup \small Fig. \thefigure: }{
 \endgroup
 \end{minipage}
 \end{center}}

\documentstyle[preprint,aps,eqsecnum,epsf]{revtex}
\begin{document}
\draft

\title{Innermost Stable Circular Orbit of a Spinning Particle\\
in\\
 Kerr Spacetime}
\author{{\sc Shingo} SUZUKI\thanks{electronic
mail:shingo@gravity.phys.waseda.ac.jp} and 
{\sc Kei-ichi}
 MAEDA\thanks{electronic mail:maeda@mn.waseda.ac.jp}}
\address{Department of Physics, Waseda University,
Shinjuku-ku, Tokyo 169, Japan}
\date{\today}
\maketitle

\begin{abstract}
We study stability of a circular orbit of a spinning test
particle in a Kerr spacetime. We find that 
some of the circular orbits become unstable  in the direction
perpendicular to the equatorial plane, although the orbits are
still stable in the radial direction.
Then for the large spin case (${\cal S}$ \mbox{\raisebox{-1.ex}{$\stackrel{\textstyle<}
{\textstyle \sim}$}} O(1)), the innermost stable
circular orbit (ISCO) appears before the minimum of the effective
potential in the equatorial plane disappears. This changes the radius of ISCO and then the
frequency of the last circular orbit. 
\end{abstract}

\section{Introduction}
\label{sec1}\setcounter{equation}{0}
Coalescence of two compact objects, such as neutron star-neutron 
star, neutron star-black hole and/or black hole-black hole
 binaries, is one of the promising sources of gravitational
waves  which may be detected in the near future by a laser
interferometric  detector, such as US LIGO \cite{abromovici}. If
we  detect a signal of gravitational waves emitted from
those systems and compare it with theoretical templates, we may
be able not only to determine a variety of astrophysical
parameters of the sources, e.g., their orbital information,
masses, and  spins, but also to obtain more information about
fundamental physics, e.g., the
equation of state at high density\cite{cutler}. In order to
obtain such important information from the observed data, we
have to make an exact template and then we need to know the exact
motion of the binary. Hence the equations of motion in
post-Newtonian framework have been studied  by many
authors\cite{damour}. We can see  from the obtained equations of 
motion that spins of stars plays a very important role.
For example, this effect induces a precession of the orbital
plane,   resulting in modulation of the gravitational wave forms
\cite{kidder}.
Although such  post-Newtonian approaches are definitely
important, a fully general relativistic treatment must become
necessary at some stage.   Numerical relativity is one of most
promising treatments to give a correct template, but it is
still at a developing stage.  Instead, a black hole perturbation
technique has been  well studied and provides us a good
approximation.  Then we expect that a general
relativistic spin effect is also well understood by such a
perturbation approach.

Many studies about a relativistic spin effect using a spinning 
test particle around a black hole have been made since the basic
equations were derived first by Papapetrou\cite{papa} and
reformulated by Dixon\cite{dixon}. Bailey and Israel
elaborated this system using the Lagrangian formalism\cite{Bailey}.
Corinaldesi and Papapetrou first discussed the motion of a
spinning test particle in Schwarzschild spacetime\cite{papa2}.
The Kerr or Kerr-Newman spacetime case was also analyzed by
several authors\cite{Rasband}. 
 In \cite{mino}, the gravitational waves produced by 
a spinning particle falling into a Kerr black hole
 or moving circularly around it were discussed and the energy 
emission rate from those systems was calculated.  We 
investigated more generic motion of a spinning particle around a
Schwarzschild black hole and pointed out that 
the spin effect can make some orbits  chaotic\cite{shingo}.

When we discuss the gravitational waves from a binary system,
one of the most important keys is the innermost stable circular
orbit (ISCO). In the binary system,
as gravitational waves are emitted, both energy and angular
momentum of the system  decrease, and so the orbital
radius gradually decreases.  When the angular momentum is
reduced below some critical value, we cannot find any circular
orbit. The binary system evolves into a more dynamical stage
from the quasi-stationary stage. The last circular orbit is the
ISCO. It may determine the observable frequency of the system.
Using the effective
potential of the spinning particle in the equatorial plane, the
binding energy of the circular orbit is discussed in
\cite{Rasband}.  They assumed that a
stable circular orbit exists unless the minimum point of the
effective potential in the equatorial plane disappears, finding
the largest binding energy.

   However, before the system reaches such a last orbit, if there exists any unstable mode, it
will change the ISCO, and therefore the last circular radius, binding energy and frequency,
which may be important in gravitational wave astronomy. In fact, we found some circular orbits
of a spinning particle in Schwarzschild spacetime  become unstable in the direction
perpendicular to the equatorial plane\cite{shingo}.

In this paper, extending our previous analysis\cite{shingo}, we
study an instability of the spinning test particle in a Kerr
spacetime. The reason for a transition from the quasi-periodic stage to
dynamical stage, we have so far three effects. 
Most important one is general relativistic strong gravity, i.e. when the orbital
separation of the binary becomes small as 
$r\sim 6M$, the centrifugal force
 cannot be balanced with the gravitational force, so that 
the  star cannot stop  from changing  its circular orbit 
to a plunging one. 
The second effect is the
hydrodynamical interaction between two stars, i.e. at the moment
when the surfaces of the stars
 come into contact, we expect that such a transition will occur
due to the direct interaction between stars.
 The third one is an effect due to a property of the star such as a spin or a multipole
moment. For example, in
\cite{Lai1}, they show that the instability can occur due to a tidal force.  When the orbital
separation of the stars becomes small,  each star of the binary is significantly deformed by a
tidal force.  As a result, a circular orbit of the stars becomes unstable and
the ISCO appears earlier. 
In this
analysis, however, the motion of the stars is restricted into the 
equatorial plane and  the stars plunge to each other when the
motion becomes unstable in the radial direction. Here, we look for
 a new type of instability due to an effect belongs to the third category,  i.e. the
instability
 in the direction perpendicular to the equatorial plane. 
We will show that such an instability really occurs
when the spin of a test particle is high. 
A spinning particle leaves the equatorial plane and falls
into black hole. 
Since it is important to determine the ISCO for the gravitational wave
astronomy, we believe that it is also worthwhile to analyze such a
new instability of a spinning particle in Kerr spacetime.

 This paper is organized as follows. After a brief review of the
basic equations in section 2, we analyze  stability  of  a
circular orbit of a spinning test particle in section 3, using
a linear perturbation method. We show changes of  the radius of
 the ISCO, of the binding energy and
of the frequency due to such an instability.
 Summary and some remarks follow in section 4.
 Throughout this paper we use units $c=G=1$.  Our notation including the 
signature of the metric follows MTW\cite{gravitation}.
\section{Basic Equations for a Spinning Test Particle}
\label{sec2}\setcounter{equation}{0}
\subsection{Pole-dipole Approximation}
The equations of motion of a spinning test particle in a
relativistic spacetime were first derived by  Papapetrou\cite{papa}
and then reformulated by Dixon\cite{dixon}.
These are a set of equations:
\begin{eqnarray}
\frac{dx^{\mu}}{ds}&=&v^{\mu},
\label{eqn:xdot}
\\
\frac{Dp^{\mu}}{ds}&=&
-\frac{1}{2}R^{\mu}_{~\nu\rho\sigma}v^{\nu}S^{\rho\sigma},
\label{eqn:pdot}
\\
\frac{DS^{\mu\nu}}{ds}&=&p^{\mu}v^{\nu}-p^{\nu}v^{\mu},
\label{eqn:sdot}
\end{eqnarray}
where $s, v^{\mu}, p^{\mu}$ and $S^{\mu\nu}$ are an affine
parameter of the orbit $x^{\mu}=x^{\mu}(s)$, the 4-velocity of the
particle, the momentum, and the spin tensor, respectively. This is called the pole-dipole
approximation, where the multipole moments of the particle higher than
mass monopole  and spin dipole are ignored.  We need  a supplementary
condition which gives a  relation between
$v^{\mu}$ and  $p^{\mu}$, because $p^{\mu}$ is no longer parallel
to $v^{\mu}$. The consistent choice of the center of mass
provides such a condition\cite{dixon},
\begin{equation}
p_{\mu}S^{\mu\nu}=0.
\label{eqn:ps0}
\end{equation}
  Using (\ref{eqn:ps0}) we find the
relation between $v^{\mu}$ and $p^{\mu}$ as
\begin{equation}
v^{\mu}=N\left[u^{\mu}
+\frac{1}{2\mu^2\delta}S^{\mu\nu}u^{\lambda} 
R_{\nu\lambda\rho\sigma}
S^{\rho\sigma}\right],
\label{eqn:p-v}
\end{equation}
where 
\begin{equation}
\delta=1+\frac{1}{4\mu^2}R_{\alpha\beta\gamma\delta}
S^{\alpha\beta}S^{\gamma\delta},
\end{equation}
and $N$ is a normalization constant, which is fixed by a choice of the
affine parameter $s$.
 $u^{\nu} \equiv p^{\nu}/\mu$ is a normalized momentum, where the mass
of the particle $\mu$ is defined by 
\begin{equation}
\mu^2=-p_{\nu}p^{\nu}.
\label{eqn:mass}
\end{equation}

It may be sometimes more convenient or more intuitive to describe 
the basic equations by use of a spin vector $S_\mu$, which is
defined by
\begin{equation}
S_\mu = - \frac{1}{2}\epsilon_{\mu\nu\rho\sigma}
\label{eqn:svector}
u^{\nu}S^{\rho\sigma},
\label{def:spinvec}
\end{equation}
where $\epsilon^{\mu\nu\rho\sigma}$ is the Levi-Civita tensor.
The basic equations are now
\begin{eqnarray}
\frac{dx^{\mu}}{ds}&=&v^{\mu},\\
\frac{Dp^{\mu}}{ds}&=&
\frac{1}{\mu}R^{*\mu}_{~~\nu\rho\sigma}v^{\nu}S^{\rho}p^{\sigma},\\
\frac{DS^{\mu}}{ds}&=&
\frac{1}{\mu^3}p^{\mu}R^{*}_{~\nu\lambda\rho\sigma}S^{\nu}v^{\lambda}S^{\rho}p^{\sigma},
\end{eqnarray}
where
\begin{equation}
R^{*}_{~\mu\nu\rho\sigma}\equiv\frac{1}{2}
R_{\mu\nu}^{~~\alpha\beta}\epsilon_{\alpha\beta\rho\sigma}.
\end{equation}
Eq. (\ref{eqn:ps0}) with the definition (\ref{eqn:svector}) reads
\begin{equation}
p_{\mu}S^{\mu}=0,
\label{eqn:ps0vec}
\end{equation}
which gives the relation between $v^{\mu}$ and $p^{\mu}$,
\begin{equation}
v^{\mu}=u^{\mu}+\frac{1}{\mu^2}{}^{*}\!R^{*\mu}_{~~\nu\rho\sigma}
S^{\nu}S^{\rho}u^{\sigma},
\label{eqn:p-vvec}
\end{equation}
where
\begin{equation}
{}^{*}\!R^{*}_{~\mu\nu\rho\sigma}\equiv\frac{1}{2}
\epsilon_{\mu\nu\alpha\beta}R^{*\alpha\beta}_{~~~\rho\sigma}.
\end{equation}
by fixing the affine parameter $s$ using the condition
$v^{\mu}u_{\mu}=-\delta$. This choice makes the perturbation equations
simpler as we will show later. 

For a particle motion in a Kerr spacetime, we find several conserved
quantities. Regardless of the symmetry of the background spacetime, it
is  easy to show that $\mu$  and the
magnitude of spin ${\cal S}$, defined by  
\begin{equation}
{\cal S}^2\equiv S_{\mu}S^{\mu},
\label{eqn:spin}
\end{equation}
are  constants of motion.
When the spacetime possesses some symmetry described by a  Killing 
vector $\xi^{\mu}$,
\begin{equation}
C\equiv\xi^{\mu}p_{\mu}-\frac{1}{2}\xi_{\mu ;\nu}S^{\mu\nu}
\label{eqn:killing}
\end{equation}
is conserved\cite{dixon}. For the spacetime which has both axial 
and timelike Killing vectors such as Kerr
spacetime, we have two conserved quantities, i.e. the energy $E$ and the $z$
component of the total angular momentum of a
spinning particle $J_z$. For Schwarzschild spacetime, the
``$x$'' and ``$y$'' components of the total angular momentum of a
particle are also conserved
 because of spherical symmetry.
\subsection{Effective potential of a spinning particle on the
equatorial plane} 
In order to discuss the motion of a test particle in a black hole
spacetime, we usually introduce an effective potential. However, if
the test particle has a spin, it is not so easy to find an effective
potential because of an additional dynamical freedom of the spin
direction. In our previous paper, we defined an ``effective potential''
for a spinning test particle in  Schwarzschild spacetime, i.e.
$V_{(\pm)}(r,\theta;J,{\cal S})$  given by Eqs.(2.31) and (2.32) in
\cite{shingo}. Setting the total angular momentum as 
 to point in the $z$ direction, i.e. $(J_x,J_y,J_z)=(0,0,J)$, those equations are obtained
from the condition of
$p^r=p^\theta=0$ in (\ref{eqn:ps0}) and (\ref{eqn:mass}). We use quotation marks because it
is not a real effective potential but plays a similar role.

The particle with energy $E$ can move within the contour curve defined
by
\begin{equation}
E=V_{(\pm)}(r,\theta;J,{\cal S}).
\end{equation}
This ``effective potential'' has two parameters, $J$ and ${
\cal S}$, whose values determine the topology of its contours. From the contours of the
``effective potential'' (Fig.2 in \cite{shingo}), we can easily see the stability of a
bound orbit.

However, since we are interest in a Kerr background spacetime, whose metric is given
as
\begin{equation}
ds^2=-\left(1-\frac{2Mr}{\Sigma}\right)dt^2
+\frac{\Sigma}{\Delta}dr^2+\Sigma d\theta^2+\frac{{\cal
A}}{\Sigma}\sin^2\theta d\phi^2-\frac{4Ma}{\Sigma}r\sin^2\theta
dtd\phi,
\end{equation}
where,
\begin{equation}
\Sigma=r^2+a^2\cos^2\theta,\;\;\;\Delta = r^2-2Mr+a^2,\;\;\;{\cal
A}=(r^2+a^2)^2-a^2\Delta\sin^2\theta,
\end{equation}
it is hard to define an effective potential even in the above
naive sense. This is because we cannot find new conserved
quantities in a Kerr background spacetime, which correspond to $J_x$, $J_y$ for Schwarzschild
case or Carter's constant for a spinless particle\cite{gravitation}, in addition to $E$ and
$J_z$. However, if we restrict the particle motion in the
equatorial plane, an effective potential for the radial motion is found\cite{Rasband}.
The direction of spin must be parallel to the rotational axis. In
fact, setting $\theta=\pi/2$, $p^2=0$, $S^0=S^1=S^3=0$ and $S^2 \equiv
-{\cal S}$, we find that Eq. (\ref{eqn:mass}) is reduced to,
\begin{equation}
(p_{1})^2=A(E-U_{(+)})(E-U_{(-)}),
\label{eqn:eff_kerr}
\end{equation}
and
\begin{equation}
U_{(\pm)}(r;J_z,{\cal S},a)=XJ_z\pm\sqrt{(X^2-Y)J_z^2-Z},
\label{eqn:potentialKerr}
\end{equation}
\begin{eqnarray}
X&=&\frac{\left[r^2+a^2+\frac{a{\cal
S}}{\mu}\left(1+\frac{M}{r}\right)\right]\left(a+\frac{M{\cal S}}{\mu
r}\right)-\Delta\left(a+\frac{{\cal
S}}{\mu}\right)}{\left[r^2+a^2+\frac{a{\cal
S}}{\mu}\left(1+\frac{M}{r}\right)\right]^2-\Delta\left(a+\frac{S}{\mu}\right)^2},\nonumber\\
Y&=&\frac{\left(a+\frac{M{\cal S}}{\mu r}\right)^2-\Delta}{\left[r^2+a^2+\frac{a{\cal
S}}{\mu}\left(1+\frac{M}{r}\right)\right]^2-\Delta\left(a+\frac{{\cal
S}}{\mu}\right)^2},\nonumber\\
Z&=&-\frac{\Delta\left(\frac{M{\cal S}^2}{\mu^2r^2}-r\right)\mu^2}{\left[r^2+a^2+\frac{a{\cal
S}}{\mu}\left(1+\frac{M}{r}\right)\right]^2-\Delta\left(a+\frac{{\cal
S}}{\mu}\right)^2},\nonumber\\
A&=&\frac{\left[r^2+a^2+\frac{a{\cal
S}}{\mu}\left(a+\frac{M}{r}\right)\right]^2-\Delta\left(a+\frac{{\cal
S}}{\mu}\right)^2}{\Delta\left(\frac{M{\cal S}^2}{\mu^2 r^2}-r\right)^2}.\nonumber
\end{eqnarray}
Note that the suffices of the momentum and the spin vector denote those tetrad components. The
tetrad frame has been defined as
\begin{eqnarray}
e^{0}_{\mu}&=&\left(\sqrt{\frac{\Delta}{\Sigma}},0,0,
-a\sin^2\theta\sqrt{\frac{\Delta}{\Sigma}}\right)\nonumber,\\
e^{1}_{\mu}&=&\left(0,\sqrt{\frac{\Sigma}{\Delta}},0,0\right),\\
e^{2}_{\mu}&=&\left(0,0,\sqrt{\Sigma},0\right)\nonumber,\\
e^{3}_{\mu}&=&\left(-\frac{a}{\sqrt{\Sigma}}\sin\theta,0,0,
\frac{r^2+a^2}{\sqrt{\Sigma}}\sin\theta\right)\nonumber,
\end{eqnarray}
where $e^{a}_{\mu}=(e^{a}_{t},e^{a}_{r},e^{a}_{\theta},
e^{a}_{\phi})$ for $a=0\sim 3$. We use of ${\cal S}$ including the direction of
spin as well as the magnitude. Then $a{\cal S}>0$ means the spin of the particle is parallel to
that of the black hole and $a{\cal S}<0$ means anti-parallel.

We regard $U_{(\pm)}(r;J_z,{\cal S},a)$ as an effective potential 
of the particle on the equatorial plane.  When $a$ vanishes,
$U_{(+)}$ is reduced to $V(r,\pi/2;J,{\cal S})$ in the
Schwarzschild case\cite{shingo}. The particle with energy $E$ can
move only in the region of $E\geq U_{(+)}$ (or $E\leq U_{(-)}$) on the
equatorial plane. The typical shape of $U_{(\pm)}$ is shown in Fig.1. 
As for the orbit in the equatorial plane, we usually discuss it by use of this effective
potential. In the case of a spinless particle, it is enough because the angular momentum
conservation prevent the particle from leaving off the equatorial plane. If the angular
momentum is larger than some critical value, we always find a stable circular orbit. Since a
circular orbit does not exist below the critical angular momentum, the particle
eventually enters a dynamical stage from a quasi-periodic stage because the emitted
gravitational waves bring the angular momentum of the system away. This critical value gives
the ISCO. If a test particle has a spin, however, the orbital angular momentum needs not be
conserved. The particle may leave off the equatorial plane. Hence the analysis by use of the
above effective potential may not give the ISCO. We need a more detailed analysis to see
whether the circular orbit on the equatorial plane is really stable or not. 
\singlefig{13cm}{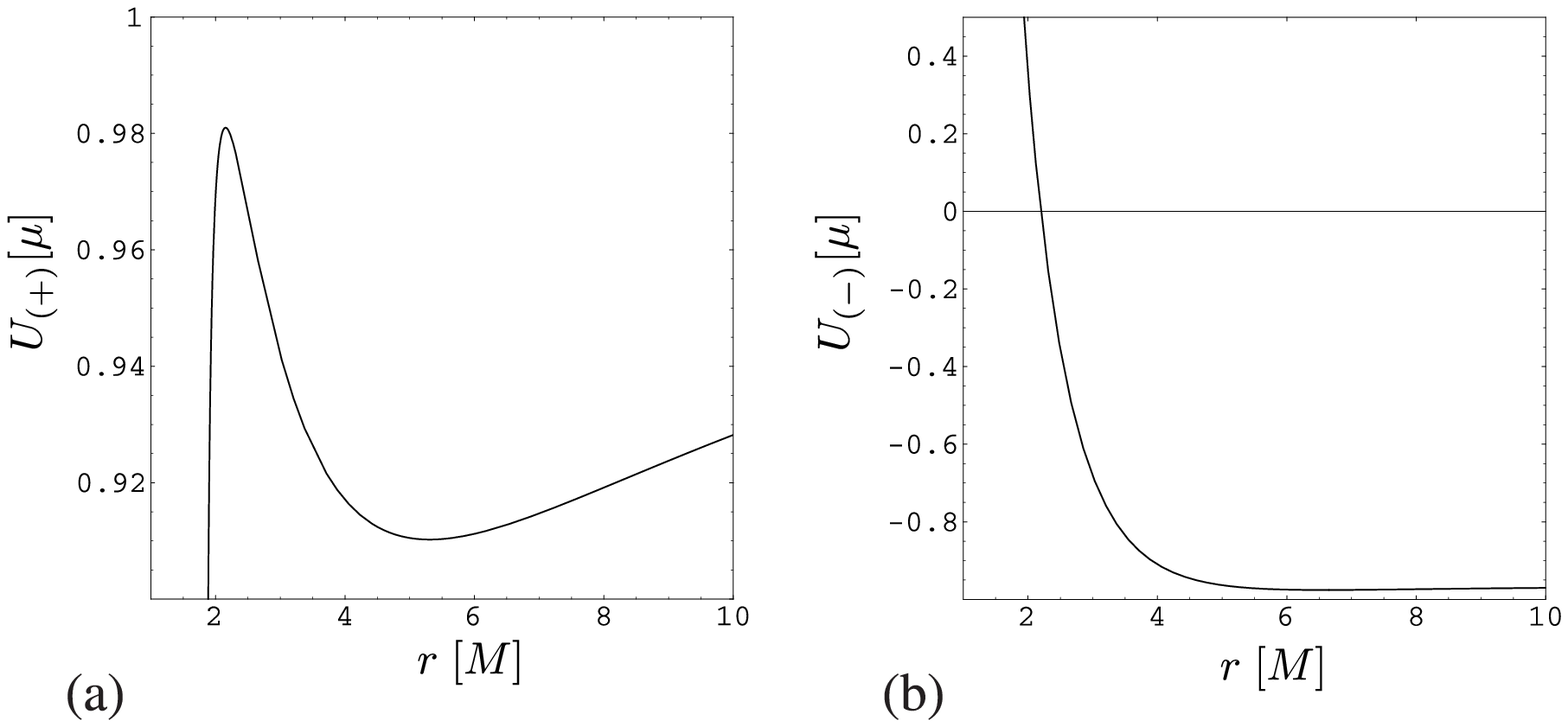}
\begin{figcaption}{fig01}{14.5cm}
The effective potential $U_{(\pm)}$
on the equatorial plane for $J_z=4 \mu M,\;{\cal S}=1 \mu M$ and 
$a=0.5 M$.  (a) The potential $U_{(+)}$ has two extremal points. (b) The potential $U_{(-)}$
has no extremum. A particle with positive energy is not bounded. 
\end{figcaption}
\section{Stability of a Circular Orbit of a Spinning Particle}
\label{sec3}\setcounter{equation}{0}
The linear perturbation method is used to analyze the stability 
of a circular orbit of a spinning particle.
In the tetrad frame, the equations of the motion of a spinning 
particle are rewritten as
\begin{eqnarray}
\frac{dx^{\mu}}{ds}&=&e^{\mu}_{a}v^{a},\label{eqn:eqs_tetrad1}\\
\frac{dp^{a}}{ds}&=&\omega_{cb}^{~~a}v^{c}p^{b}
+\frac{1}{\mu}R^{*a}_{~~bcd}v^{b}S^{c}p^{d},\\
\frac{dS^{a}}{ds}&=&\omega_{cb}^{~~a}v^{c}S^{b}
+\frac{1}{\mu^3}p^{a}S^{i}R^{*}_{~ijkl}v^{j}S^{k}p^{l},
\label{eqn:eqs_tetrad3}
\end{eqnarray}
where $\omega_{ijk}$ are the Ricci rotation coefficients defined as
\begin{equation}
\omega_{ijk}=e^{\nu}_{i}e^{\mu}_{j}e_{k\mu;\nu}.
\end{equation}

We assume a circular orbit in the equatorial plane as an unperturbed orbit,  i.e.
$r_{(0)}=r_0=\mbox{const.},\;\theta_{(0)}=\pi/2,\;p^{1}_{(0)}=p^{2}_{(0)}=0,\;
S^{0}_{(0)}=S^{1}_{(0)}=S^{3}_{(0)}=0,\;S^{2}_{(0)}=-{\cal S}$ and the time derivatives of
these variables vanish. Here, we use the suffix
$(0)$ for the unperturbed variables. From Eq.(\ref{eqn:p-vvec}), we find
\begin{eqnarray}
v^{0}_{(0)}&=&\frac{p^{0}_{(0)}}{\mu}\left(1-\frac{M{\cal
S}^2}{\mu^2r^3}\right),\\
v^{3}_{(0)}&=&\frac{p^{3}_{(0)}}{\mu}\left(1+\frac{2M{\cal S}^2}{\mu^2r^3}\right).
\end{eqnarray}
and $v^{1}_{(0)}=v^{2}_{(0)}=0$.
 The conserved quantities in our system are the energy and
the
$z$ component of total angular momentum, which are now given as
\begin{eqnarray}
E&=&\frac{\sqrt{\Delta}}{r_0}p^0_{(0)}+\frac{1}{r_{0}}\left(a+\frac{M{\cal
S}}{\mu}\right)p^3_{(0)},\label{eqn:E}\\
J_z&=&\frac{\sqrt{\Delta}}{r_0}\left(a+\frac{{\cal
S}}{\mu}\right)p^0_{(0)}+\frac{1}{r_0}\left(r_0^2+a^2 +\frac{a{\cal
S}}{\mu}\left(1+\frac{M}{r_0}\right)\right)p^3_{(0)},\label{eqn:Jz}
\end{eqnarray}
The way to determine the unperturbed variables is as follows.
First, the Kerr parameter $a$ and the magnitude of the spin 
${\cal S}$ are fixed. Next, we give the radius of a circular
orbit $r_0$. Then, from the condition of potential extremum at $r_0$,
i.e. 
\begin{equation}
\left.\frac{dV_{(+)}(r)}{dr}\right|_{r=r_0}=0,
\end{equation}
$J_z$ is determined. The energy of the particle is obtained by
$E=V_{(+)}(r_0)$.
Finally, from Eqs.(\ref{eqn:E}) and (\ref{eqn:Jz}), $p^{0}_{(0)}$ and
$p^{3}_{(0)}$ are determined. We denote these unperturbed variables as
$x^{\mu}_{(0)},\;v^{i}_{(0)},\;p^{i}_{(0)}$ and $S^{i}_{(0)}$. Of course, if no real solution
$J_z$ is found for given parameters $a$ and ${\cal S}$, a circular orbit does not exist at
$r_0$.

Let $\delta x^{\mu},\;\delta v^{i},\;\delta p^{i}$ and $\delta S^{i}$
be the perturbations around the unperturbed variables. We shall introduce a tetrad description
for $\delta x^{\mu}$ as 
$\delta x^{i}\equiv e^{i}_{\mu}\delta x^{\mu}$. From Eq. (\ref{eqn:p-vvec}), 
\begin{equation}
\delta v^{i}=X^{i}_{j}\delta x^{j} +Y^{i}_{j}\delta
p^{j}+Z^{i}_{j}\delta S^{j},
\end{equation}
where,
\begin{eqnarray}
X^{i}_{~j}&=&
\frac{1}{\mu}e^{\mu}_{j}\frac{\partial}{\partial x^{\mu}}{}^{*}\!R^{*~~i}_{(0)~lmn}
S^{l}_{(0)}S^{m}_{(0)}p^{n}_{(0)},\label{eqn:Xij}\\
Y^{i}_{~j}&=&
\frac{1}{\mu}\delta^{i}_{j}+\frac{1}{\mu^3}{}^{*}\!
R^{*~~i}_{(0)~lmn}S^{l}_{(0)}S^{m}_{(0)}\delta^{n}_{j},\\
Z^{i}_{~j}&=&
\frac{1}{\mu^3}{}^{*}\!R^{*~~i}_{(0)~lmn}(\delta^{l}_{j}S^{m}_{(0)}
+S^{l}_{(0)}\delta^{m}_{j})p^{n}_{(0)}.
\end{eqnarray}
Inserting
\begin{eqnarray}
x^{\mu}&=&x^{\mu}_{(0)}+e^{\mu}_{i}\delta x^{i},\nonumber\\
v^{i}&=&v^{i}_{(0)}+\delta v^{i},\\
p^{i}&=&p^{i}_{(0)}+\delta p^{i},\nonumber\\
S^{i}&=&S^{i}_{(0)}+\delta S^{i},\nonumber
\end{eqnarray}
into Eqs.(\ref{eqn:eqs_tetrad1})-(\ref{eqn:eqs_tetrad3})  and
neglecting higher terms than the first order of perturbed variables, we find the
linear perturbation equations as
\begin{equation}
\frac{d}{ds}\left(
\begin{array}{c}
\delta x^{i}\\
\delta p^{i}\\
\delta S^{i}
\end{array}
\right)=\left(
\begin{array}{ccc}
A_{~j}^{i}&B_{~j}^{i}&C_{~j}^{i}\\
D_{~j}^{i}&E_{~j}^{i}&F_{~j}^{i}\\
G_{~j}^{i}&H_{~j}^{i}&I_{~j}^{i}
\end{array}
\right)
\left(
\begin{array}{c}
\delta x^{j}\\
\delta p^{j}\\
\delta S^{j}
\end{array}
\right),
\label{eqn:perturbation}
\end{equation}
where
\begin{eqnarray}
A_{~j}^{i}&=&X^{i}_{~j}+\omega_{(0)lk}^{~~~~~i}\delta^{k}_{j}v^{l}_{(0)}
-\omega_{(0)lk}^{~~~~~i}v^{k}_{(0)}\delta^{l}_{j}\nonumber,\\
B_{~j}^{i}&=&Y^{i}_{~j}\nonumber,\\ 
C_{~j}^{i}&=&Z^{i}_{~j}\nonumber,\\
D_{~j}^{i}&=&e^{\mu}_{j}\frac{\partial}{\partial
x^{\mu}}\omega_{(0)lk}^{~~~~~i}v^{l}_{(0)}p^{k}_{(0)}
+\omega_{(0)lk}^{~~~~~i}X^{l}_{~j}p^{k}_{(0)}+\Phi^{i}_{~j}\nonumber,\\
E_{~j}^{i}&=&\omega_{(0)lk}^{~~~~~i}v^{l}_{(0)}\delta^{k}_{j}
+\omega_{(0)lk}^{~~~~~i}Y^{l}_{~j}p^{k}_{(0)}+\Pi^{i}_{~j},\\ \label{eqn:component}
F_{~j}^{i}&=&\omega_{(0)lk}^{~~~~~i}Z^{l}_{~j}p^{k}_{(0)}+\Psi^{i}_{~j}\nonumber,\\
G_{~j}^{i}&=&e^{\mu}_{j}\frac{\partial}{\partial
x^{\mu}}\omega_{(0)lk}^{~~~~~i}v^{l}_{(0)}S^{k}_{(0)}
+\omega_{(0)lk}^{~~~~~i}X^{l}_{~j}S^{k}_{(0)}+\frac{1}{\mu^2}p^{i}_{(0)}S_{(0)l}\Phi^{l}_{~j},\nonumber\\
H_{~j}^{i}&=&\frac{1}{\mu^3}R^{*}_{(0)abcd}S^{a}_{(0)}v^{b}_{(0)}S^{c}_{(0)}p^{d}_{(0)}\delta^{i}_{j}
+\omega_{(0)lk}^{~~~~~i}Y^{l}_{~j}S^{k}_{(0)}
+\frac{1}{\mu^2}p^{i}S_{(0)l}\Pi^{l}_{~j}\nonumber,\\
I_{~j}^{i}&=&\omega_{(0)lj}^{~~~~~i}v^{l}_{(0)}
+\frac{1}{\mu^3}p^{i}_{(0)}R^{*}_{(0)jabc}v^{a}_{(0)}S^{b}_{(0)}p^{c}_{(0)}
+\omega_{(0)lk}^{~~~~~i}Z^{l}_{~j}S^{k}_{(0)}
+\frac{1}{\mu^2}p^{i}_{(0)}S_{(0)l}\Psi^{l}_{~j}\nonumber,
\end{eqnarray}
with
\begin{eqnarray}
\Phi^{i}_{~j}&=&\frac{1}{\mu}\left(e^{\mu}_{j}\frac{\partial}{\partial
x^{\mu}}R^{*~~i}_{(0)~lmn}v^{l}_{(0)}
+R^{*~~i}_{(0)~lmn}X^{l}_{~j}\right)S^{m}_{(0)}p^{n}_{(0)}\nonumber,\\
\Pi^{i}_{~j}&=&\frac{1}{\mu}R^{*~~i}_{(0)~lmn}(Y^{l}_{~j}p^{n}_{(0)}
+v^{l}_{(0)}\delta^{n}_{j})S^{m}_{(0)},\\
\Psi^{i}_{~j}&=&\frac{1}{\mu}R^{*~~i}_{(0)~lmn}(Z^{l}_{~j}S^{m}_{(0)}
+v^{l}_{(0)}\delta^{m}_{j})p^{n}_{(0)}\nonumber.
\end{eqnarray}
Eq.(\ref{eqn:perturbation}) is nothing but the equation of 
geodesic deviation for the case of a spinless particle. The eigenvalues of the matrix in
Eq.(\ref{eqn:perturbation}) determine the stability of a test particle.
Inserting the unperturbed variables into the matrix in Eq.(\ref{eqn:perturbation}), the
matrix becomes rather simple.
In Appendix, we show the explicit form the matrix and its components and discuss the structure
of the matrix.
The eigenequation forms
\begin{equation}
\lambda^6(\lambda^2-\Lambda_r)(\lambda^4 -\Lambda_{\theta
1}\lambda^2+\Lambda_{\theta 2})=0,
\label{eqn:eigeneq}
\end{equation}
where $\Lambda_{r},\;\Lambda_{\theta 1}$ and $\Lambda_{\theta 2}$ are defined in Appendix.
First, to see the meaning of  this
equation,  we consider the simple case .  When ${\cal S}=0$ and $a=0$, this equation is reduced
to
\begin{equation}
\lambda^8\left(\lambda^2
+\frac{M\mu^2(r_0-6M)}{r_0^3(r_0-3M)}\right)\left(\lambda^2
+\frac{L_z^2}{r_0^4}\right)=0,
\label{eqn:eigen_S0}
\end{equation}
i.e. 
\begin{eqnarray}
\Lambda_r&=&-\frac{M\mu^2(r_0-6M)}{r_0^3(r_0-3M)},\nonumber\\
\Lambda_{\theta 1}&=&-\frac{L_z^2}{r_0^4},\\
\Lambda_{\theta 2}&=&0,\nonumber
\end{eqnarray}
where $L_z$ is the $z$ component of orbital angular  momentum, and it is given for an
unperturbed circular orbit as
\begin{equation}
L_{z}^2=\frac{M\mu^2r_0^2}{r_0-3M}.
\end{equation}
This equation shows that any circular  orbit
does not exist for $r_0<3M$.  The positivity of the second parenthesis of (\ref{eqn:eigen_S0})
guarantees the stability in the $\theta$ direction, i.e. in the direction perpendicular to the
equatorial plane. Then, the orbit of a
spinless particle in Schwarzschild spacetime is stable against perturbations in the
$\theta$ direction.  The first parenthesis in (\ref{eqn:eigen_S0}) shows the stability in
the $r$ direction. For $r_0\leq6M$, the eigenvalues are real, while
 for $r>6M$, those are pure imaginary. This means that the equilibrium point (circular orbit)
with 
$r_0\leq 6M$ is unstable, while it is stable for the case of $r_0>6M$. We find the
critical radius
$r=6M$ when $L_z=\sqrt{12}\mu M$.  This corresponds to the ISCO of
a spinless particle in Schwarzschild spacetime.

In the general case, we can analyze stability of the circular orbit in the $r$ direction
by $\Lambda_{r}$ and  in the
$\theta$ direction by $\Lambda_{\theta 1}$ and $\Lambda_{\theta
2}$ (see also Appendix). If $\Lambda_r>0$, the circular orbit is unstable against a radial
perturbation, while if a real part of any solutions of the equation
\begin{equation}
\lambda^4-\Lambda_{\theta 1}\lambda^2+\Lambda_{\theta 2}=0,
\end{equation}
is positive, it is unstable against
perturbations perpendicular to the equatorial plane.

In Fig.2, we show the unstable parameter ranges by shaded regions for the case of 
$a=0,\;0.4M$ and $0.8M$. We find the unstable parameter range is usually narrow, because we
have a constraint for a particle spin as ${\cal S}/\mu M$
\mbox{\raisebox{-1.ex}{$\stackrel{\textstyle<} {\textstyle \sim}$}} O(1) (see discussion in
\cite{shingo}). Although the generic features of the parameter range is similar, there is some
small difference depending on $a$. As we  can see from this figure, the unstable region in the
$a=0.4M$ case is  larger than that in the $a=0$ case. The value of ${\cal S}$ at the bottom
edge of this region is $\sim 0.9\mu M$. However, in the $a=0.8M$ case, the unstable region
becomes narrow again. The lowest value of ${\cal S}$ for this region is about $0.94\mu M$.

Fig.3 shows the radius of the ISCO for several values of $a$. 
The radius of the ISCO decreases as ${\cal S}$ increases or as $a$ 
increases, which means that the spin effect plays the role of a
repulsive force for $a{\cal S}>0$ basically. Therefore, the
particle can approach the horizon of the black hole enduring the
gravity. But, if ${\cal S}$ is larger than the critical value, the
radius of the ISCO increases. This shows the occurence of the
instability of the motion in the $\theta$ direction. In Fig.4, we show the energy of the
particle and the orbital frequency of the ISCO. In \cite{Rasband}, the energy
the ISCO for same system was analyzed, but only the instability in the radial direction was
taken into account. When the instability in the $\theta$ direction is included, the
energy and frequency of the ISCO behaves in different manner beyond the critical value. 
\singlefig{13.5cm}{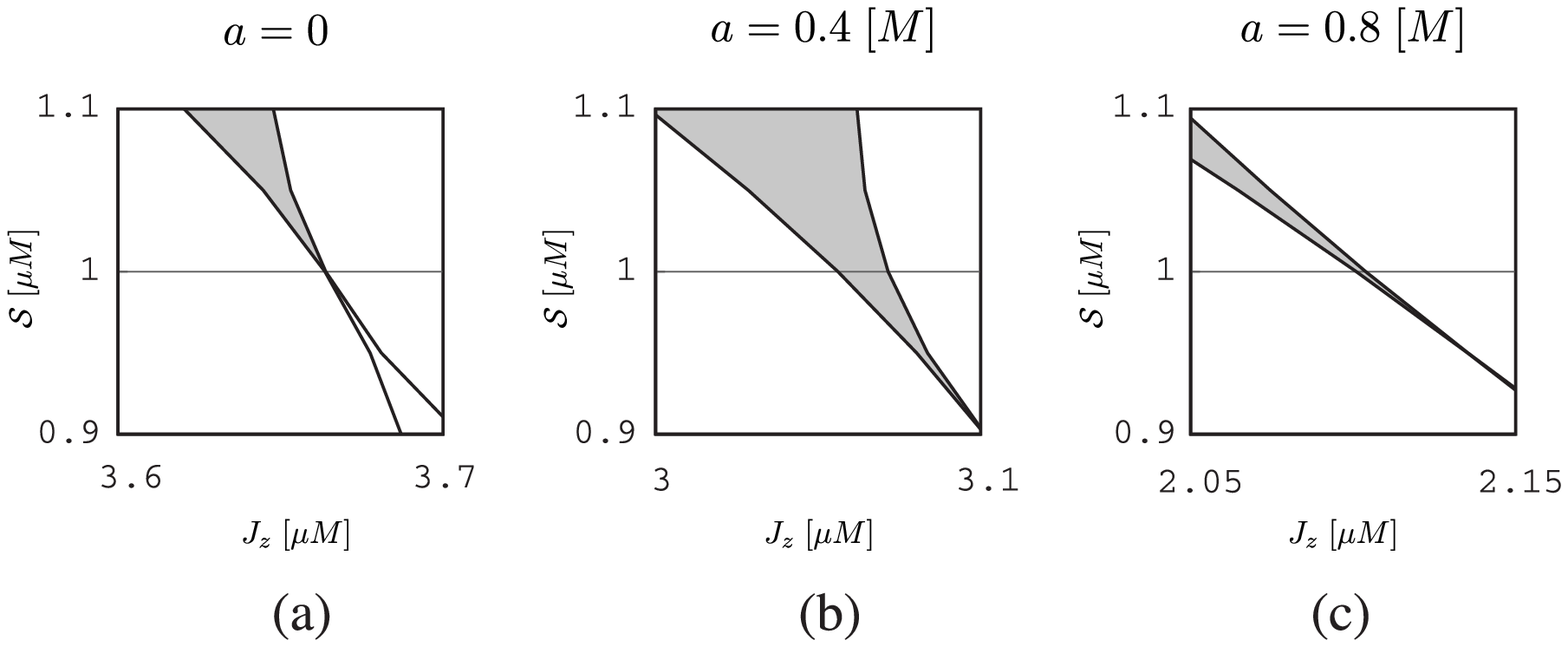}
\begin{figcaption}{fig02}{14.5cm}
The parameter regions for which the circular orbit becomes unstable in the direction
perpendicular to the equatorial plane. The minimal values of the spin for each case are (a)
${\cal S}=0.98\mu M$ (b)
${\cal S}=0.9\mu M$ and (c)
${\cal S}=0.94\mu M$ 
\end{figcaption}
\singlefig{9cm}{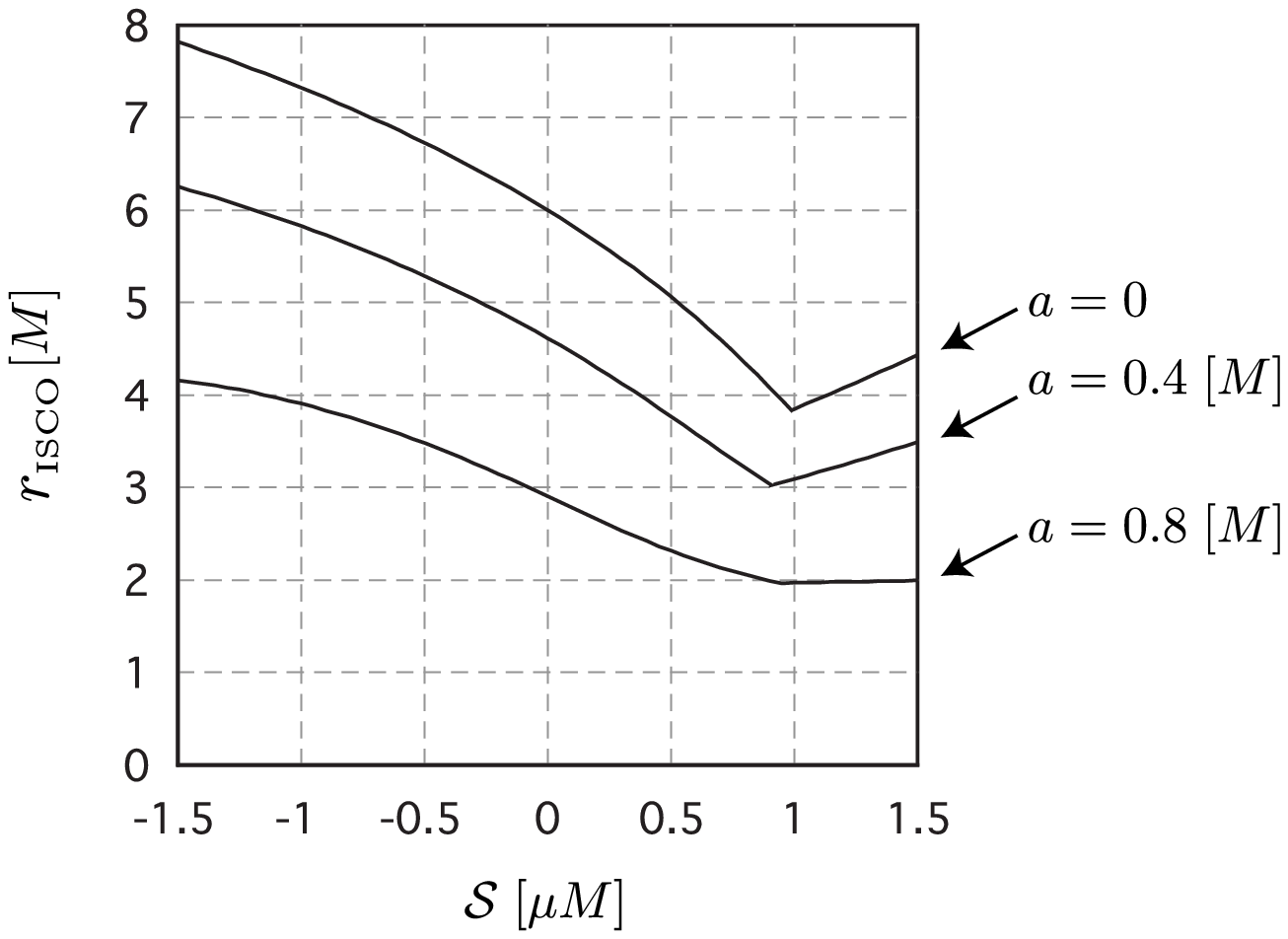}
\begin{figcaption}{fig03}{14.5cm}
The radius of the innermost stable circular orbit (ISCO). Above the critical value of ${\cal
S}$, the radius of the ISCO increases.
\end{figcaption}
\singlefig{13.5cm}{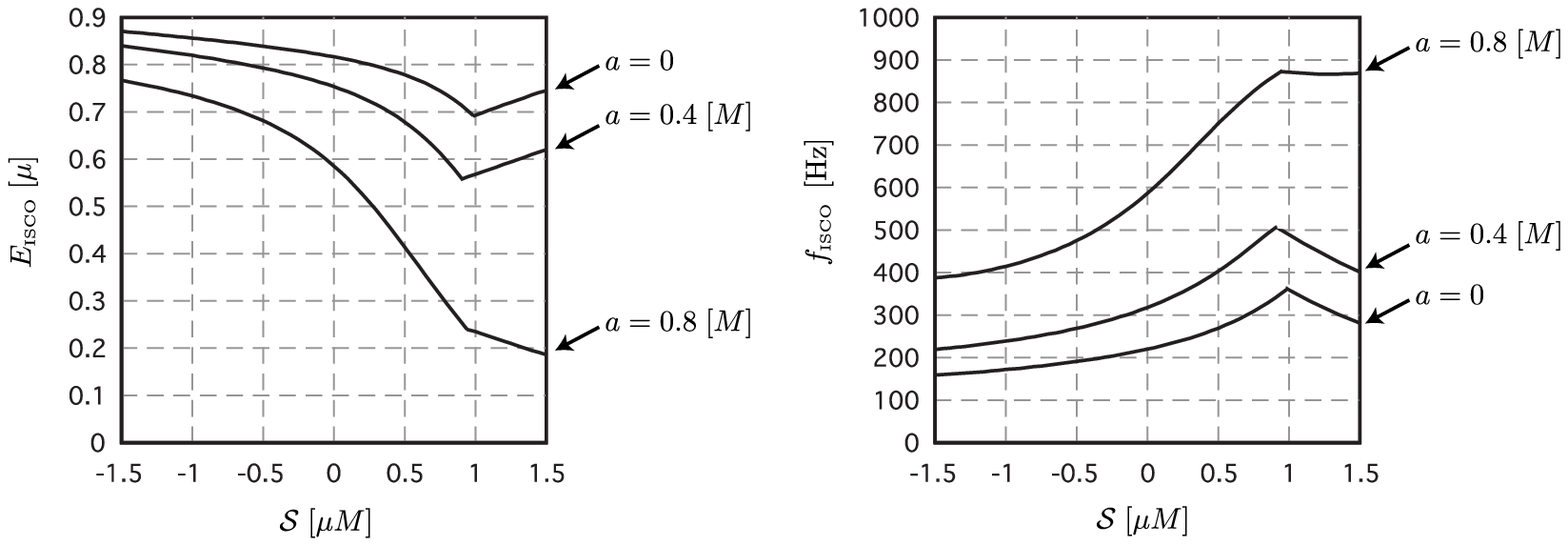}
\begin{figcaption}{fig04}{14.5cm}
(a) The particle energy $E$ of the ISCO. (b) The orbital frequency of the ISCO for which the 
mass of the black hole is assumed to be $10M_{\odot}$ 
\end{figcaption}
\section{ Discussions and Summary}
\label{sec5}\setcounter{equation}{0}
In our previous paper\cite{shingo}, we discussed the motion of a spinning test particle in a
Schwarzschild spacetime by use of an ``effective potential". The ``effective potential" is
classified into four types depending on its topology of contour (Figs.2 and 3 in
\cite{shingo}). From Fig.2 in \cite{shingo}, we can see easily the stability of the circular
orbit of the spinning particle in Schwarzschild background spacetime. The type(B1) has one
saddle point and one minimal point on the equatorial plane. The saddle point is maximal in the
$r$ direction and minimal in the
$\theta$ direction. At the minimal point, it is minimal in both directions. For
the type(B2) potential, there are two saddle points apart from the equatorial plane and a
maximal and minimal points also exist on the equatorial plane. In the type(U2) potential,
we find one unstable point in both directions and one saddle point. At this saddle point, we
find that a particle is stable in the $r$ direction but unstable in the
$\theta$ direction. This is exactly the  same as what we found in this paper as new type of
instability. The stability of the circular orbit on the equatorial plane is summarized in Table
1. 

For a Kerr background spacetime, however, we cannot define the effective potential for a
spinning particle like Schwarzschild case. Thereupon, instead of the topology of the contour,
we divided the parameter space by the stability of the circular orbit. The result for $a=0$
case is shown in Fig.6. The name of classified types in this figure are same as those of the
``effective potential'' in \cite{shingo} in which the circular orbit shows the same
stability type as the present analysis. Since the region of $J_z\geq 3.5\mu M$ and ${\cal
S}\geq 0$ coincides with Fig.3 in \cite{shingo}, we believe that the present classification
is the same as the previous one by use of the ``effective potential''. Fig.7 is the
classification for the Kerr background spacetime. We expected that the parameter region for
type(U2)  potential for Kerr background spacetime would be larger than that for
Schwarzschild  spacetime because of the lower symmetry and the existence of the spin-spin 
interaction. However, as we have seen in Fig.2, the parameter region is not enlarged
abruptly as $a$ increases. Rather, for large
$a$, this region gets narrow.

Since our system is not integrable as shown in \cite{shingo}, it may be interesting to
mention chaos of a spinning test particle in Kerr background spacetime.
In our previous paper, we conclude that the motion of the particle
can be chaotic in the type(B2) potential. As we see the type(B2) region becomes larger as $a$
increases. Therefore, we expect that the motion of the particle in Kerr spacetime may become
chaotic more easily than in Schwarzschild spacetime. This is consistent with the fact that
Kerr spacetime possesses lower symmetry than Schwarzschild spacetime and that number of the
constants of motion is smaller than that in Schwarzschild case. As for a example, we show in
Fig.8 the Poincar\'e map of the chaotic orbit for the
$a=0.8M$ case constructed with the same rule in
\cite{shingo}. The parameters of the orbit are chosen to
$J_z=4\mu M$ and ${\cal S}=0.3\mu M$, which are in the type(B2) region. For Schwarzschild
case, the magnitude of the spin at the bottom edge of the type(B2) region with $J_z=4\mu M$,
i.e. the lowest value of the magnitude spin for which the chaos occurs, is about $0.635\mu M$
(Fig.7 in \cite{shingo}), which is mush higher than that of the above example (${\cal
S}=0.3\mu m$). This result supports our expectation\cite{footnote}.

In this paper, we have analyzed the stability of a circular orbit of a spinning test particle
in a Kerr background spacetime using linear perturbation method. The circular orbit can be
unstable not only in the $r$ direction but also in the $\theta$ direction due to the spin.
With the parameters within the shaded region in Fig.2, this instability occurs. In
Fig.5, a typical unstable orbit with these parameters and its fate are shown. We have
chosen the parameters for the initially circular orbit of the particle as ${\cal S}=1\mu
M,\;E=0.86199950863\mu,\;J_z=3.06620962243\mu M,\;r_0=3.05M$ and $\theta=\pi/2$. The Kerr
parameter is $a=0.4M$. As for a perturbation, the direction of the spin is set to be
slightly different from the perpendicular to the equatorial plane, i.e. $\delta
S^3=3.672835\times 10^{-5}\mu M$. Because this circular orbit is unstable in the $\theta$
direction, the particle leaves the equatorial plane and falls into the black hole eventually.
Then this type of instability may change the radius of the ISCO. In Fig.3, the
dependence the radius of the ISCO upon the magnitude of the spin is shown. In Fig.4, the
energy of the particle on the ISCO and the orbital frequency of the ISCO are also
depicted. 

If this new type of instability occurs in a real astrophysical  system
such as a binary system, it changes the radius of the ISCO. This is very important for
observation, especially, gravitational astronomy as mentioned in Introduction.
So, we may restrict the parameters of the
system, for example the  lower limit of the spin, from the observation of gravitational waves.
And the chaos is also important factor for the gravitational wave astronomy, because
if the chaos occurs in the system, it must affect the gravitational wave emitted from
the system. However, the test particle model should not be adopted as a model  of the last
stage of a binary system. In fact, in our analysis, the radius of the ISCO gets very small.
For such small distances to the horizon, instabilities due to another effects mentioned in
Introduction must be taken into account. Therefore, we need further investigation to check the
importance of this type of instability
 in a more realistic binary system.
\newpage
\singlefig{7cm}{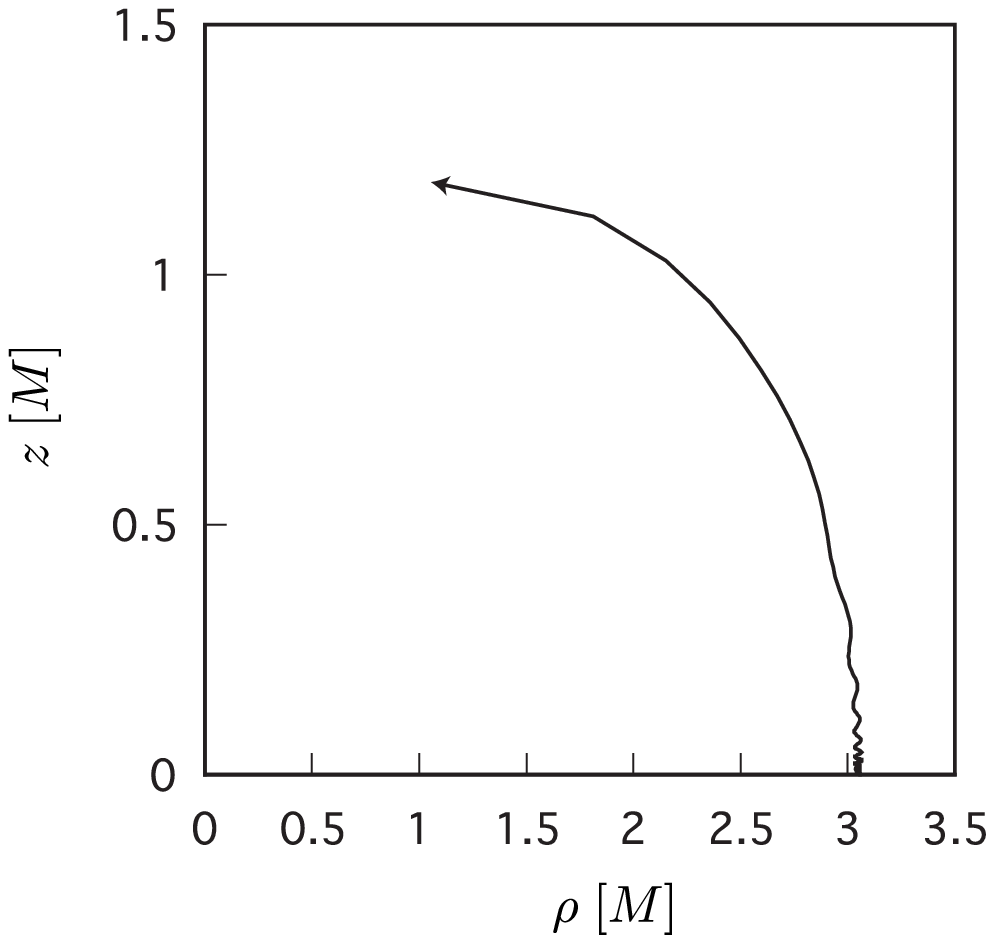}
\begin{figcaption}{fig05}{14.5cm}
The orbit of a spinning particle in type(U2) potential in Kerr background spacetime with
$a=0.4M$. The particle is at the minimum point of the  ``effective
potential'' in the $r$ direction on the equatorial plane  and the
direction of the spin vector is slightly deviated from the direction
perpendicular to the equatorial plane initially. The parameters of the orbit are
$J_z=3.06620962243\mu M,\;E=0.86199950863\mu,\;{\cal S}=1.0\mu M$ and $r_0=3.05M$. The
initial perturbation is $\delta S^{3}=3.672835\times 10^{-5}\mu M$.  
\end{figcaption}
\newpage
\singlefig{9cm}{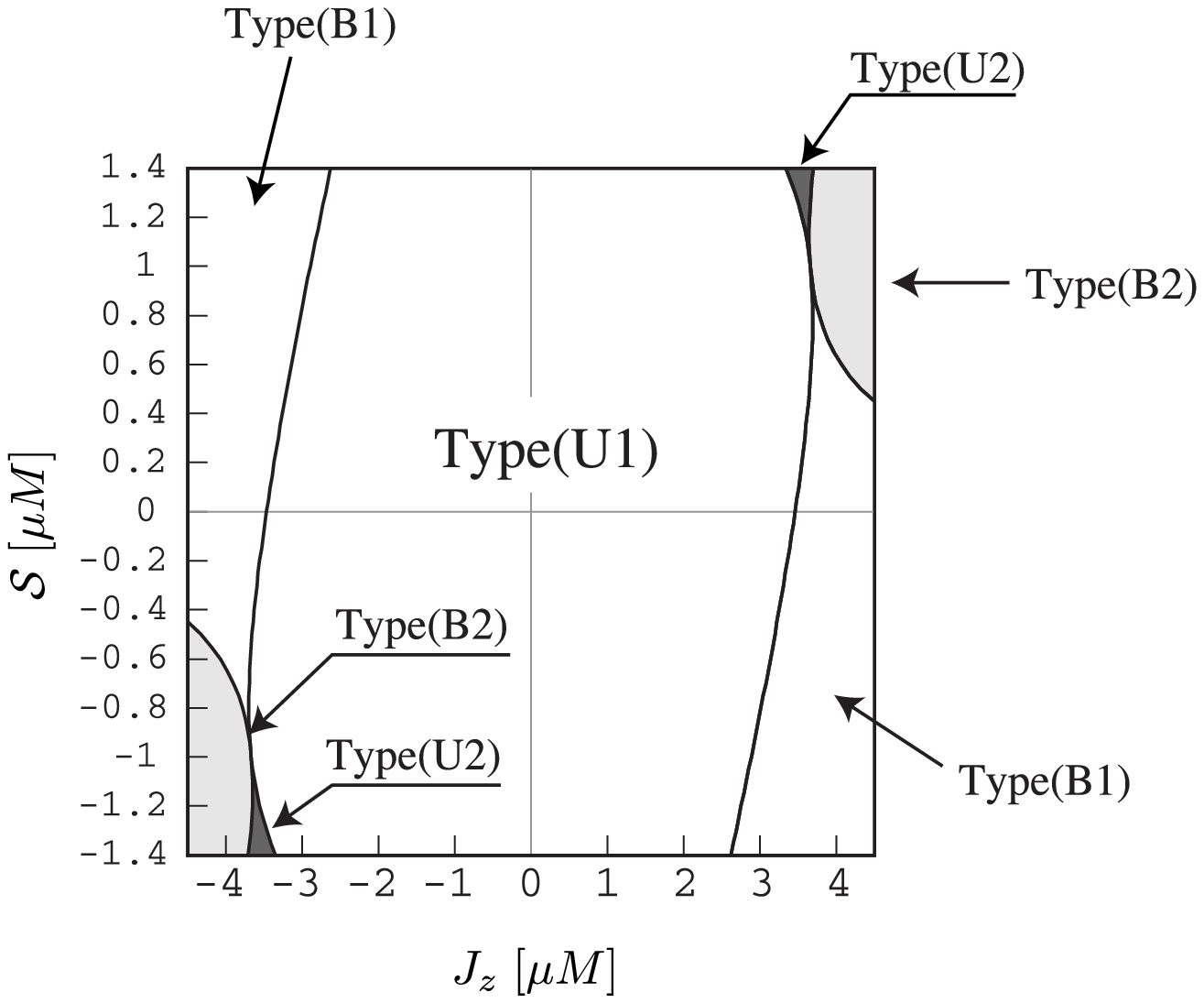}
\begin{figcaption}{fig06}{14.5cm}
The classification of the ``effective potential'' of the particle 
for the $a=0$ case in the $J_{z}$-${\cal S}$ plane. The region $J_z>3.5\mu
M$ and ${\cal S}>0$ is  the same as Fig.3 in \cite{shingo} 
\end{figcaption}
\singlefig{12cm}{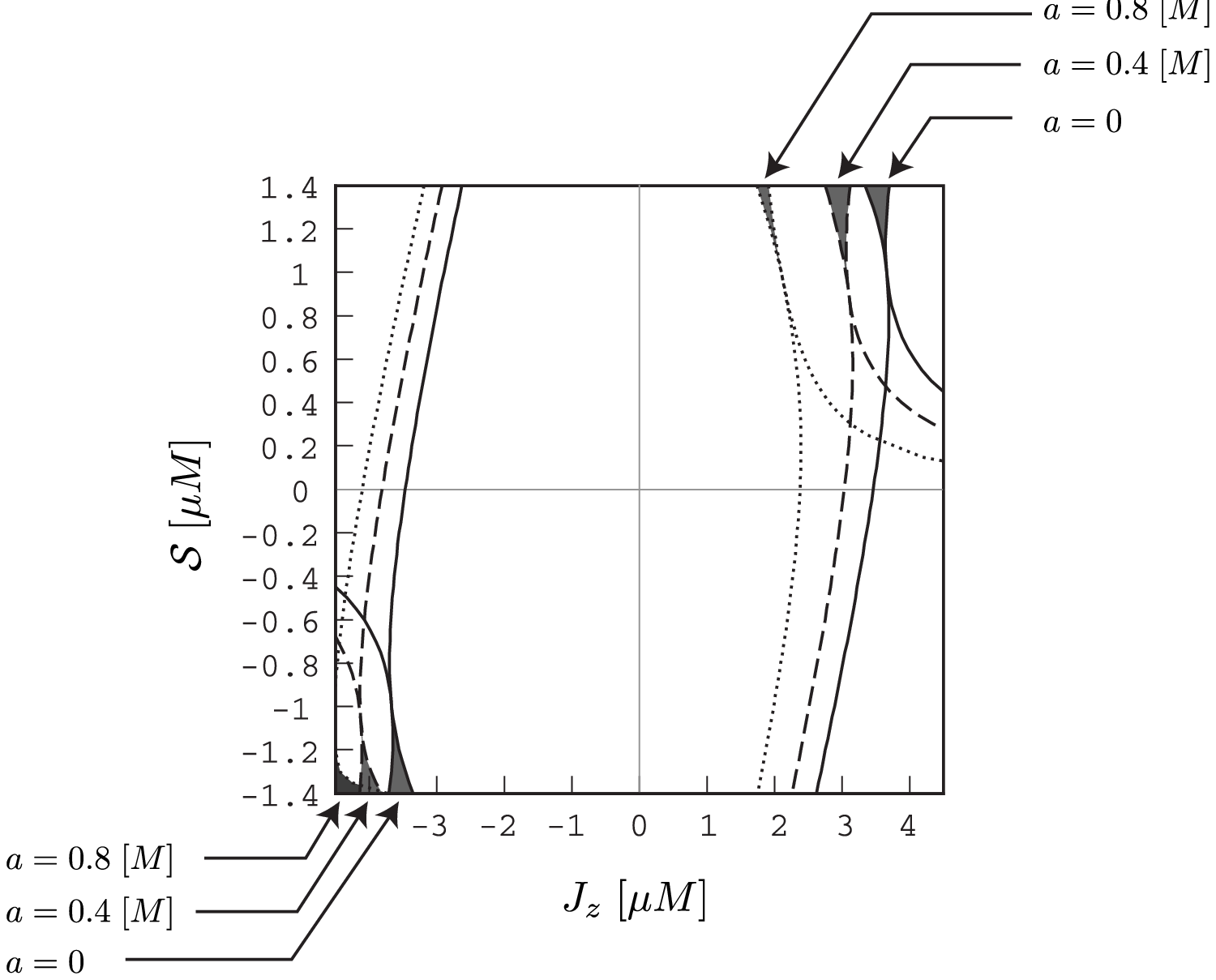}
\begin{figcaption}{fig07}{14.5cm}
The classification of the effective potential of the 
particle on the equatorial plane for a Kerr black hole. 
\end{figcaption}
\singlefig{8cm}{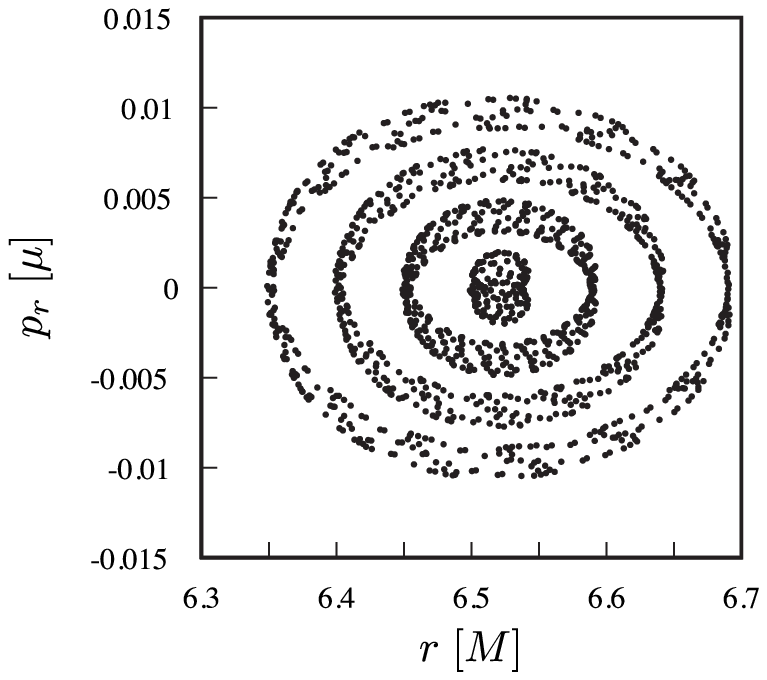}
\begin{figcaption}{fig08}{14.5cm}
The Poincar\'e map for the orbit in a Kerr    spacetime, $a=0.8M$, with parameters
which belong to the type(B2). All orbits has $J_z=4.0\mu M,\;{\cal S}=0.3\mu M$ and
$E=0.958155\mu$. We set
$p^{r}=0$ and
$S^{12}/S^{31}=\tan(85^{\circ})=11.4301$ initially. And the initial position is
$r_0=6.5,\;6.592,\;6.628$ and $6.68M$. We find the motion is chaotic.
\end{figcaption}
\vspace{2cm}
\begin{center}
\begin{tabular}{l||c|c||c|c}
\hline
&\multicolumn{2}{c||}{inner extremal point}&\multicolumn{2}{c}{outer extremal point}\\
\cline{2-5}
&$r$ direction&$\theta$ direction&$r$ direction&$\theta$ direction\\
\hline
\hline
Type(B1)&Unstable&Stable&Stable&Stable\\
\hline
Type(B2)&Unstable&Unstable&Stable&Stable\\
\hline
Type(U1)&\multicolumn{4}{c}{no extremal point}\\
\hline
Type(U2)&Unstable&Unstable&Stable&Unstable\\
\hline
\end{tabular}
\end{center}
\parbox[t]{2cm}{TABLE. 1:\\~}\ \
\parbox[t]{14cm}
{The stability of a spinning test particle on the equatorial plane. }\\[1em]
\noindent
\appendix
\section{the perturbation matrix and its eigenequation}
Here, we present the explicit form of the perturbation matrix in
Eq.(\ref{eqn:perturbation}) and its non-vanishing components.
\begin{equation}
\left(
\begin{array}{ccc}
A_{~j}^{i}&B_{~j}^{i}&C_{~j}^{i}\\
D_{~j}^{i}&E_{~j}^{i}&F_{~j}^{i}\\
G_{~j}^{i}&H_{~j}^{i}&I_{~j}^{i}
\end{array}
\right)=\left(
\begin{array}{cccccccccccc}
0&A_{~1}^{0}&0&0&B_{~0}^{0}&0&0&0&0&0&C_{~2}^{0}&0\\
0&0&0&0&0&B_{~1}^{1}&0&0&0&0&0&0\\
0&0&0&0&0&0&B_{~2}^{2}&0&C_{~0}^{2}&0&0&C_{~3}^{2}\\
0&A_{~1}^{3}&0&0&0&0&0&B_{~3}^{3}&0&0&C_{~2}^{3}&0\\
0&0&0&0&0&E_{~1}^{0}&0&0&0&0&0&0\\
0&D_{~1}^{1}&0&0&E_{~0}^{1}&0&0&E_{~3}^{1}&0&0&F_{~2}^{1}&0\\
0&0&D_{~2}^{2}&0&0&0&0&0&0&F_{~1}^{2}&0&0\\
0&0&0&0&0&E_{~1}^{3}&0&0&0&0&0&0\\
0&0&G_{~2}^{0}&0&0&0&0&0&0&I_{~1}^{0}&0&0\\
0&0&0&0&0&0&H_{~2}^{1}&0&I_{~0}^{1}&0&0&I_{~3}^{1}\\
0&0&0&0&0&0&0&0&0&0&0&0\\
0&0&G_{~2}^{3}&0&0&0&0&0&0&I_{~1}^{3}&0&0
\end{array}
\right),
\label{eqn:matrix}
\end{equation}
where
\begin{eqnarray}
A^{0}_{~1}&=&\frac{3M{\cal
S}^2\sqrt{\Delta}}{\mu^3r_0^5}p^{0}_{(0)}-\left.\left(\frac{\sqrt{\Delta}}{r}\right)^{\prime}\right|_{r=r_0}v^{0}_{(0)},\nonumber\\
A^{3}_{~1}&=&-\frac{2a}{r_0^2}v^{0}_{(0)}-\frac{\sqrt{\Delta}}{r_0^2}v^3_{(0)}-\frac{6M{\cal
S}^2\sqrt{\Delta}}{\mu^3r_0^5}p^3_{(0)},\nonumber\\
B^{0}_{~0}&=&\frac{1}{\mu}\left(1-\frac{M{\cal S}^2}{\mu^2r_0^3}\right),\nonumber\\
B^{1}_{~1}&=&\frac{1}{\mu}\left(1-\frac{M{\cal S}^2}{\mu^2r_0^3}\right),\nonumber\\
B^{2}_{~2}&=&\frac{1}{\mu},\nonumber\\
B^{3}_{~3}&=&\frac{1}{\mu}\left(1+\frac{2M{\cal S}^2}{\mu^2r_0^3}\right),\nonumber\\
C^{0}_{~2}&=&\frac{2M{\cal S}}{\mu^3r_0^3}p^{0}_{(0)},\nonumber\\
C^{2}_{~0}&=&\frac{M{\cal S}}{\mu^3r_0^3}p^{0}_{(0)},\nonumber\\
C^{2}_{~3}&=&\frac{2M{\cal S}}{\mu^3r_0^3}p^{3}_{(0)},\nonumber\\
C^{3}_{~2}&=&-\frac{4M{\cal S}}{\mu^3r_0^3}p^{3}_{(0)},\nonumber\\
D^{1}_{~1}&=&-\frac{3M{\cal
S}^2\sqrt{\Delta}}{\mu^3r_0^5}\left.\left(\frac{\sqrt{\Delta}}{r}\right)^{\prime}\right|_{r=r_0}\left(p^0_{(0)}\right)^2-\frac{3Ma{\cal
S}^2\sqrt{\Delta}}{\mu^3r_0^7}p^0_{(0)}p^3_{(0)}-\frac{6M{\cal
S}^2\Delta}{\mu^3r_0^7}\left(p^3_{(0)}\right)^2\nonumber\\
&&-\frac{\sqrt{\Delta}}{r_0}\left.\left(\frac{\sqrt{\Delta}}{r^2}\right)^{\prime\prime}\right|_{r=r_0}v^0_{(0)}p^0_{(0)}+\frac{\sqrt{\Delta}}{r_0}\left.\left(\frac{\sqrt{\Delta}}{r^2}\right)^{\prime}\right|_{r=r_0}v^3_{(0)}p^3_{(0)}\nonumber\\
&&-\frac{\sqrt{\Delta}}{\mu r_0^5}\left[(2a\mu r_0+3M{\cal
S})v^3_{(0)}p^0_{(0)}+2(a\mu r_0+3M{\cal S})v^0_{(0)}p^3_{(0)}\right],\nonumber\\
D^{2}_{~2}&=&-\frac{a}{\mu r_0^5}(a\mu r_0+3M{\cal S})v^0_{(0)}p^0_{(0)}-\frac{1}{\mu
r_0^5}[\mu r_0(r_0^2+a^2)+6M{\cal S}]v^3_{(0)}p^3_{(0)}\nonumber\\
&&-\frac{a\sqrt{\Delta}}{r_0^4}\left(v^3_{(0)}p^0_{(0)}+v^0_{(0)}p^3_{(0)}\right),\nonumber\\
E^{0}_{~1}&=&\frac{a\mu+2M{\cal S}}{\mu^2 r_0^3}\left(1-\frac{M{\cal
S}^2}{\mu^2r_0^3}\right)p^3_{(0)}-\left.\left(\frac{\sqrt{\Delta}}{r}\right)^{\prime}\right|_{r=r_0}v^0_{(0)}+\frac{a\mu+M{\cal
S}}{\mu r_0^3}v^3_{(0)},\nonumber\\
E^{1}_{~0}&=&\frac{a\mu+2M{\cal S}}{\mu^2 r_0^3}\left(1-\frac{M{\cal S}^2}{\mu^2
r_0^3}\right)p^3_{(0)}-2\left.\left(\frac{\sqrt{\Delta}}{r}\right)^{\prime}\right|_{r=r_0}v^0_{(0)}+\frac{a\mu+M{\cal
S}}{\mu r_0^3}v^3_{(0)},\nonumber\\
E^{1}_{~2}&=&\frac{a\mu+M{\cal S}}{\mu^2r_0^3}\left(1+\frac{2M{\cal S}^2}{\mu^2
r_0^3}\right)p^0_{(0)}+\frac{a\mu+2M{\cal S}}{\mu
r_0^3}v^0_{(0)}+\frac{2\sqrt{\Delta}}{r_0^2}v^3_{(0)},\nonumber\\
E^{3}_{~1}&=&-\frac{\sqrt{\Delta}}{r_0^2}v^3_{(0)},\nonumber\\
F^{1}_{~2}&=&-\frac{2M{\cal S}}{\mu^3
r_0^3}\left.\left(\frac{\sqrt{\Delta}}{r}\right)^{\prime}\right|_{r=r_0}\left(p^0_{(0)}\right)^2-\frac{2Ma{\cal
S}}{\mu^3r_0^5}p^3_{(0)}p^0_{(0)}-\frac{4M{\cal S}\sqrt{\Delta}}{\mu^3
r_0^5}\left(p^3_{(0)}\right)^2\nonumber\\
&&-\frac{M}{\mu r_0^3}\left(v^3_{(0)}p^0_{(0)}+2v^0_{(0)}p^3_{(0)}\right),\nonumber\\
G^{0}_{~2}&=&\frac{3Ma{\cal S}^2}{\mu^3
r_0^5}\left(v^0_{(0)}p^0_{(0)}+v^3_{(0)}p^3_{(0)}\right)p^0_{(0)}+\frac{a^2{\cal
S}}{r_0^4}v^0_{(0)}+\frac{a{\cal S}\sqrt{\Delta}}{r_0^4}v^3_{(0)},\nonumber\\
G^{3}_{~2}&=&\frac{3Ma{\cal
S}^2}{\mu^3r_0^5}\left(v^0_{(0)}p^0_{(0)}+2v^3_{(0)}p^3_{(0)}\right)p^3_{(0)}-\frac{a{\cal
S}\sqrt{\Delta}}{r_0^4}v^0_{(0)}-\frac{{\cal S}(r_0^2+a^2)}{r_0^4}v^3_{(0)},\nonumber\\
H^{1}_{~2}&=&-\frac{{\cal S}\sqrt{\Delta}}{\mu r_0^2},\nonumber\\
I^{0}_{~1}&=&\frac{3M{\cal
S}}{\mu^3r_0^3}\left(v^3_{(0)}p^0_{(0)}+v^0_{(0)}p^3_{(0)}\right)p^0_{(0)}-\left.\left(\frac{\sqrt{\Delta}}{r}\right)^{\prime}\right|_{r=r_0}v^0_{(0)}+\frac{a}{r_0^2}v^3_{(0)},\nonumber\\
I^{1}_{~0}&=&-\frac{M{\cal
S}^2\sqrt{\Delta}}{\mu^3r_0^5}p^0_{(0)}-\left.\left(\frac{\sqrt{\Delta}}{r}\right)^{\prime}\right|_{r=r_0}v^0_{(0)}+\frac{a}{r_0^2}v^3_{(0)},\nonumber\\
I^{1}_{~3}&=&-\frac{2M{\cal
S}^2\sqrt{\Delta}}{\mu^3r_0^5}p^3_{(0)}+\frac{a}{r_0^2}v^0_{(0)}+\frac{\sqrt{\Delta}}{r_0^2}v^3_{(0)},\nonumber\\
I^{3}_{~1}&=&\frac{3M{\cal
S}}{\mu^3r_0^3}\left(v^3_{(0)}p^0_{(0)}+v^0_{(0)}p^3_{(0)}\right)p^3_{(0)}-\frac{a}{r_0^2}v^0_{(0)}-\frac{\sqrt{\Delta}}{r_0^2}v^3_{(0)}.\nonumber
\end{eqnarray}
A prime denotes $d/dr$. It is easily shown that the eigenequation of
the matrix (\ref{eqn:matrix}) can decomposed as follows.
\begin{equation}
\lambda^3
\left|
\begin{array}{cccc}
-\lambda&0&0&B^{1}_{~1}\\
0&-\lambda&0&E^{0}_{~1}\\
0&0&-\lambda&E^{3}_{~1}\\
D^{1}_{~1}&E^{1}_{~0}&E^{1}_{~3}&-\lambda
\end{array}\right|\left|
\begin{array}{ccccc}
-\lambda&0&0&D^{2}_{~2}&F^{2}_{~1}\\
0&-\lambda&0&G^{0}_{~2}&I^{0}_{~1}\\
0&0&-\lambda&G^{3}_{~2}&I^{3}_{~1}\\
B^{2}_{~2}&C^{2}_{~0}&C^{2}_{~3}&-\lambda&0\\
H^{1}_{~2}&I^{1}_{~0}&I^{1}_{~3}&0&-\lambda
\end{array}\right|=0
\end{equation}
Three zero eigenvalues comes from two zero columns and a zero line in the matrix, which are
related to the facts that $t$ and $\phi$ are the cyclic coordinates of the system and that the
magnitude of the spin is conserved, respectively. The second part consists of the components
related to $\delta x^1,\;\delta p^{0},\;\delta p^3$ and $\delta p^{1}$. Because of the
conservation of the energy and the angular momentum of the particle, two eigenvalues of this
part are zero.The rest two eigenvalues describe the stability in $r$ direction. The
components of the last part are related to $\delta p^2,\;\delta S^{0},\;\delta S^{3},\;\delta
x^2$ and $\delta S^{1}$. The eigenequation of this part, except for  one zero eigenvalue which
comes from the condition (\ref{eqn:ps0vec}), turns out to be a quadratic equation of
$\lambda^2$ and determines the stability in the
$\theta$ direction. Finally, we find the eigenvalue equation as,
\begin{equation}
\lambda^6(\lambda^2-\Lambda_r)(\lambda^4 -\Lambda_{\theta
1}\lambda^2+\Lambda_{\theta 2})=0,
\end{equation}
where
\begin{eqnarray}
\Lambda_{r}&=&(B^1_{~1}D^1_{~1}+E^0_{~1}E^1_{~0}+E^1_{~3}E^3_{~1}),\\
\Lambda_{\theta 1}&=&(B^2_{~2}D^2_{~2}+C^2_{~0}G^0_{~2}+C^2_{~3}G^3_{~2}
+F^2_{~1}H^1_{~2}+I^0_{~1}I^1_{~0}+I^1_{~3}I^3_{~1}),\\
\Lambda_{\theta 2}&=&(C^2_{~0}F^2_{~1}G^0_{~2}H^1_{~2}
+C^2_{~3}F^2_{~1}G^3_{~2}H^1_{~2}-C^2_{~0}D^2_{~2}H^1_{~2}I^0_{~1}\nonumber\\
&&-B^2_{~2}F^2_{~1}G^0_{~2}I^1_{~0}+B^2_{~2}D^2_{~2}I^0_{~1}I^1_{~0}
+C^2_{~3}G^3_{~2}I^0_{~1}I^1_{~0}\nonumber\\
&&-B^2_{~2}F^2_{~1}G^3_{~2}I^1_{~3}-C^2_{~0}G^3_{~2}I^0_{~1}I^1_{~3}
-C^2_{~3}D^2_{~2}H^1_{~2}I^3_{~1}\nonumber\\
&&-C^2_{~3}G^0_{~2}I^1_{~0}I^3_{~1}+B^2_{~2}D^2_{~2}I^1_{~3}I^3_{~1}
+C^2_{~0}G^0_{~2}I^1_{~3}I^3_{~1}).
\end{eqnarray}
To analyze a stability of a circular orbit, we have to know $\Lambda_r$ for the stability in
the $r$ direction, while $\Lambda_{\theta 1}$ and $\Lambda_{\theta 2}$ for that in the
$\theta$ direction.


\begin{thebibliography}{99}
\bibitem{footnote} The detail analysis for chaos in Kerr background spacetime will be given
elsewhere.
\end{thebibliography}
\end{document}